## *The Comets of Caroline Herschel (1750-1848), Sleuth of the Skies at Slough*

Roberta J. M. Olson[1] and Jay M. Pasachoff[2]

[1]*The New-York Historical Society, New York, NY, USA*

[2]*Hopkins Observatory, Williams College, Williamstown, MA, USA*

Abstract.   In this paper, we discuss the work on comets of Caroline Herschel, the first female comet-hunter. After leaving Bath for the environs of Windsor Castle and eventually Slough, she discovered at least eight comets, five of which were reported in the *Philosophical Transactions of the Royal Society*. We consider her public image, astronomers' perceptions of her contributions, and the style of her astronomical drawings that changed with the technological developments in astronomical illustration.

### 1.   General Introduction and the Herschels at Bath

Building on the research of Michael Hoskin[i] and our book on comets and meteors in British art,[ii] we examine the comets of Caroline Herschel (1750-1848), the first female comet-hunter and the first salaried female astronomer (Figure 1), who was more famous for her work on nebulae. She and her brother William revolutionized the conception of the universe from a Newtonian one—i.e., mechanical with God as the great clockmaker watching over its movements—to a more modern view—i.e., evolutionary.

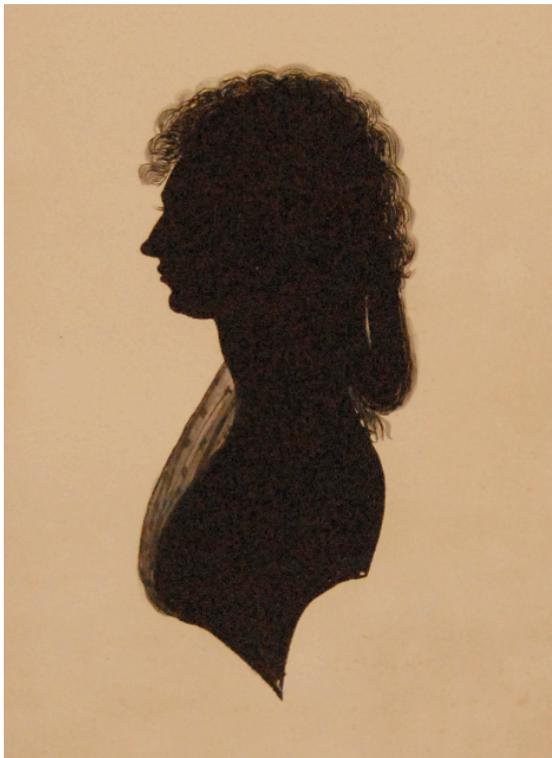

Figure 1.   Silhouette of Caroline Herschel, c. 1768, MS. Gunther 36, fol. 146r © By permission of the Oxford University Museum of the History of Science

Both Herschels began as amateur astronomers during the golden age of British astronomy, when amateurs contributed the lion's share of observations. As important transitional figures, the siblings, who both had initial careers as musicians, became among the first professional astronomers. Instead of making music—and in the case of Caroline also housekeeping—they turned to the music of the spheres.

Born Carolina Lucretia Herschel in Hanover, Germany, when England and Hanover were united under King George II, she was not physically distinguished, and her father predicted that she would die a spinster. Petite Caroline had been disfigured by smallpox—contracted at age three—and her growth stunted by typhus—caught at age eleven—so that her cold mother intended to use her as a domestic. Closer to her father, Caroline recounted that one of her strongest childhood memories was of her father taking her outside on a frosty night and showing her the winter stars 'to make me acquainted with the most beautiful constellations, after we had been gazing at a comet which was then visible'.[iii] This may, in fact, have been her defining moment, lying behind her later fame and linking her lives in Hanover and England.

After a brief German career, Caroline's favorite brother, William, began working in England in 1757, as a musician who had deserted from the Hanoverian guards, and, via an interest in mathematics and lenses, by 1773 had become obsessed with astronomy and speculative cosmology about deep space. He saved Caroline from domestic drudgery by inviting her to join him in the resort town of Bath, promising to pay their mother for a maid to assume the domestic chores! Caroline arrived at this spa—renowned for its music, theatre, and throbbing *bon ton*—on 24 August 1772, knowing only the English she had learned to parrot on the journey. In William's shadow she thrived in this elegant watering hole that drew aristocrats for the social season. After William provided her lessons in music and deportment, Caroline enjoyed a brief career as a vocalist, performing with him in Bath at the fashionable Pump Room. She also managed their home, copied music, and helped William mould and polish mirrors for telescopes. While in Bath, William had given her a telescope, a modest reflector, and introduced her to people interested in astronomy, but only with their move near Windsor Castle, would her career as an astronomer, albeit playing second fiddle, be in the ascent.

**2. Windsor: Caroline Begins her Astronomical Career and Comet-hunting**

Since astronomy was a topical subject at the time, King George III desired to entertain guests at Windsor Castle with after-dinner astronomical demonstrations. Following William's discovery of Uranus and the presentation of his reflector at Greenwich, the monarch invited William to do the honors as court astronomer with an annual pension of £200 (the Astronomer Royal earned £300). Subsequently, William named the new planet he had discovered *Georgium Sidus*. With his royal appointment, the Herschels departed Bath for the village of Datchet near Windsor, their residence from August 1782.

An aquatint by Paul Sandby (Figure 2)—based on a series of watercolors by his brother Thomas, one that is reproduced as Figure 3—recording the meteor of 18 August 1783 captures the flavor of these post-prandial viewings on the North Terrace of Windsor

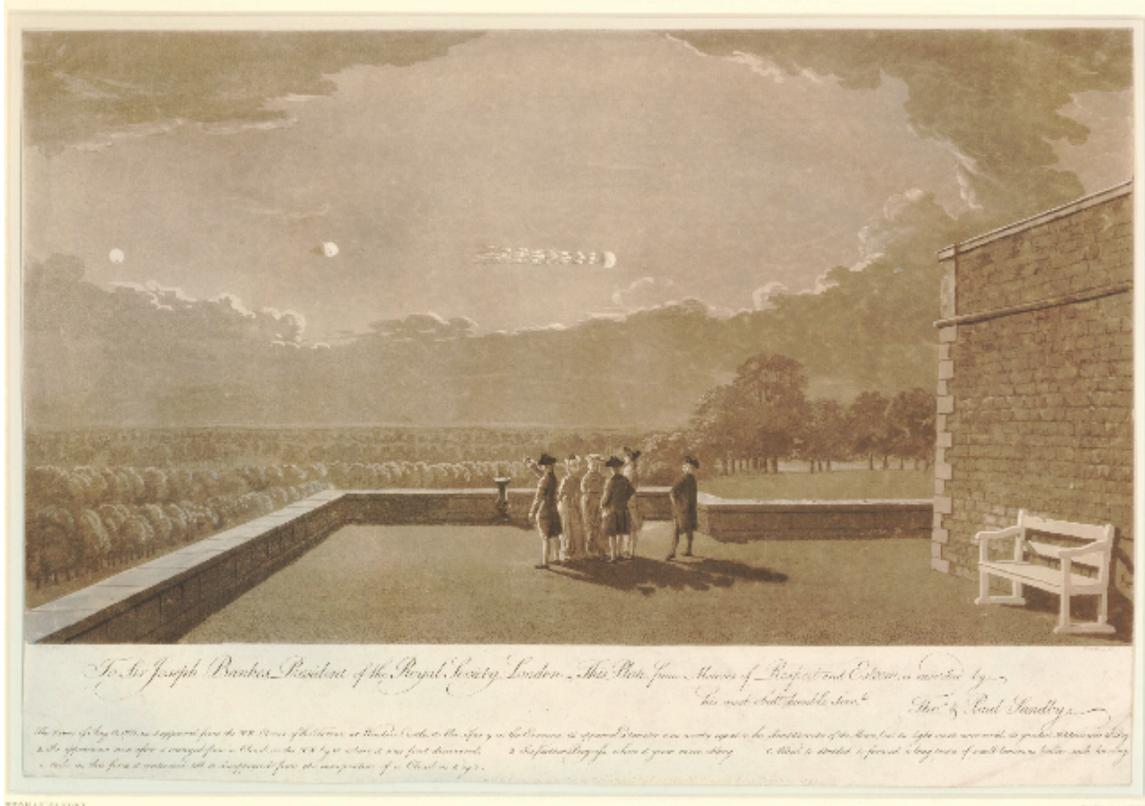

Figure 2.    Paul Sandby, The Meteor of 18 August 1783, 1783, aquatint, The British Museum, London

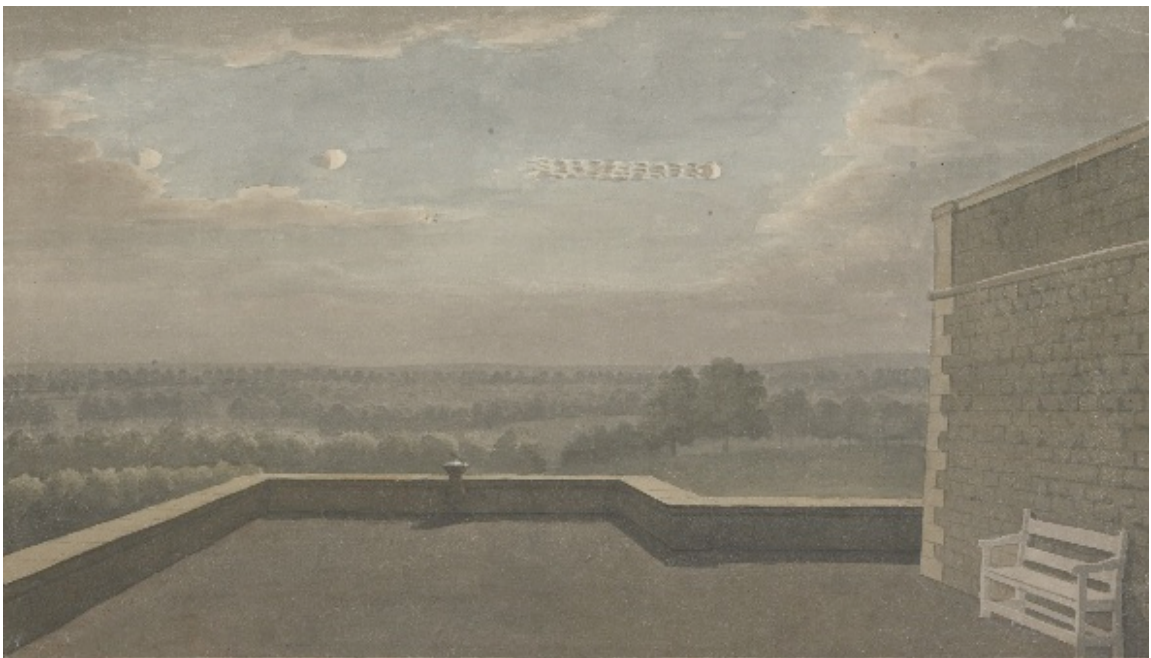

Figure 3.   Thomas or Paul Sandby, The Meteor 18 August 1783, 1783, watercolor over graphite, The British Museum, London

Castle.[iv] Thomas—who for 50 years served as deputy ranger of Windsor Great Park, architect, landscape gardener, and topographical draftsman—lived there, as did his brother Paul, and befriended both Herschels, sharing with them common interests in music first and astronomy second.[v] The rare meteor procession that the Sandbys depicted, which some people thought a comet, was a landmark in meteoric observation. Since it was bright and exploded near the end of its path, it was a bolide. Several watercolors by the brothers preserve its appearance and suggest the ambiance in which the Herschels operated, wherein astronomical observation was both for entertainment and scientific discovery. Tiberius Cavallo was one of a number of fellows of the Royal Society—including several friends of the Herschels, like Edward Piggott, Alexander Aubert, and Charles Blagden, later secretary of the Royal Society—who also observed this fireball and drew images of it that they published with their accounts in the *Philosophical Transactions of the Royal Society*. Cavallo had observed the haunting phenomenon from the castle terrace with Dr. James Lind (Caroline's physician), Dr. Lockman, Thomas Sandy, and a few others—two of whom were women. Blagden's account mentions that William Herschel watched the dazzling meteor for the longest time from his observatory at Datchet. Because Caroline was probably at his side, she is not one of the unidentified women in the Sandbys' representations, and she does not mention the meteor in the brief notes of her observations that evening.

At Datchet, William continued his observations, including searching for nebulae, one of the other astronomical *objets du jour*. Caroline, his faithful companion in star-gazing, copied down his shouted descriptions and helped him to become one of the great observers of all time. In contrast to the cosmopolitan life of Bath, the country near Windsor was lonely. To keep her occupied, over time William made her three telescopes to replace the early model he had given her in Bath, for which there is no record of her having used it. Meant as incentives, these telescopes demonstrate her growing sophistication as an observer. In August 1782, he gave her a Newtonian sweeper (Figure 4); this little refractor was inconvenient to use because it necessitated her to move around it. (The term 'sweep' is credited to William in 1786 by the Oxford English Dictionary.) Impressed with her discoveries, in 1783, William built her a small Newtonian reflector, cleverly designed for a vertical sweep of the sky simply by turning a handle. In 1788, William made her a similar five-foot focal length instrument, which she gave to her nephew, John Herschel, and today is unlocated.[vi] In June 1785, the Herschels moved to Clay Hall in Old Windsor, transferring in March 1786 to 'The Grove', later called Observatory House, at Slough (demolished in 1960), where they worked more or less until William's death in 1822. Caroline eventually lived in the adjacent cottage, which had its own observing platform on the roof.

Along with Charles Messier, dubbed by Louis XV 'the comet-ferret', Caroline Herschel was among the most assiduous observers of the skies. Like Messier, she also studied nebulae and deep sky objects to distinguish them from comets. Eventually, her assistance in sweeping for nebulae inspired William to accelerate his efforts. She was a spark as well as his muse, and finally his amanuensis. Because William could not interrupt his examination of the sky to take notes, he installed Caroline at a window to keep records with a clock and dial. When William pulled a cord as a signal, she opened the window and copied down his shouted instructions, going to bed around 4:00 a.m. The next day from her notes she wrote up her brother's nocturnal observations, which she referred to as 'minding the heavens', in her terse but lucid prose. When finished, she prepared a catalogue that the Royal Society published in 1802 in its *Philosophical Transactions* in William's name, listing around 500 new nebulae and clusters to the 2000 odd that they had previously discovered.[vii] At first, Caroline gave the term 'sweeping' a certain domestic familiarity, implying that she was a sort of celestial housekeeper, brushing and dusting the stars to keep them in a good state for a brother, a sort of heavenly *Hausfrau*.[viii]

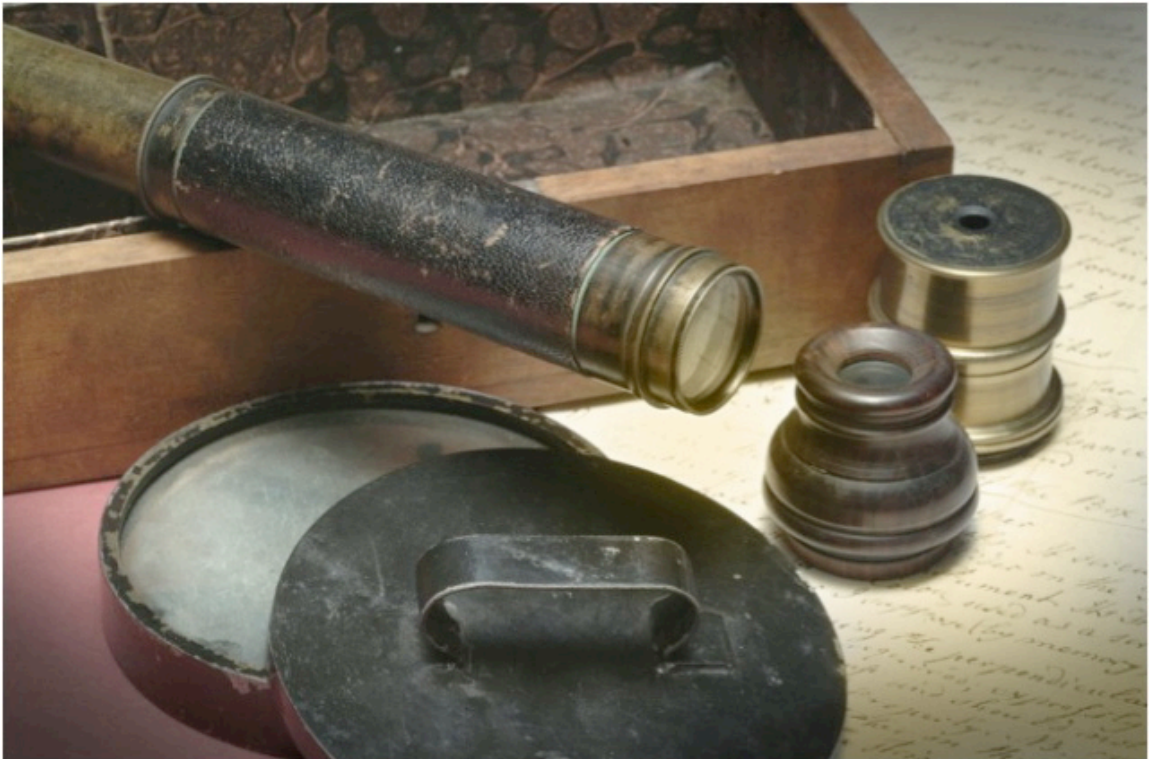

Figure 4.   Caroline's Sweeper, c. 1783, Historisches Museum, Hanover

Eventually, Caroline began to work more independently. She initiated her first record book inscribed on three opening pages: 'This is what I call the Bills & Rec.ds of my Comets', 'Comets and Letters', 'Book of Observations From Augt. 28 1782 to Feb. 5, 1787'.[ix] It, together with two subsequent books, belongs to the invaluable Herschel trove, including the papers of William and John, held by the Royal Astronomical Society in London. On another initial page, she recorded William's instructions to search for comets and nebulae. On 19 November, shortly after the meteor of 18 August 1783 had increased the awareness of celestial phenomena, Piggott discovered a comet that Caroline later spotted and recorded as 'Piggott's comet'. She observed it from 29 November through 19 December and drew it twice inside viewing fields (Figure 5).[x] William recorded it as well.[xi]

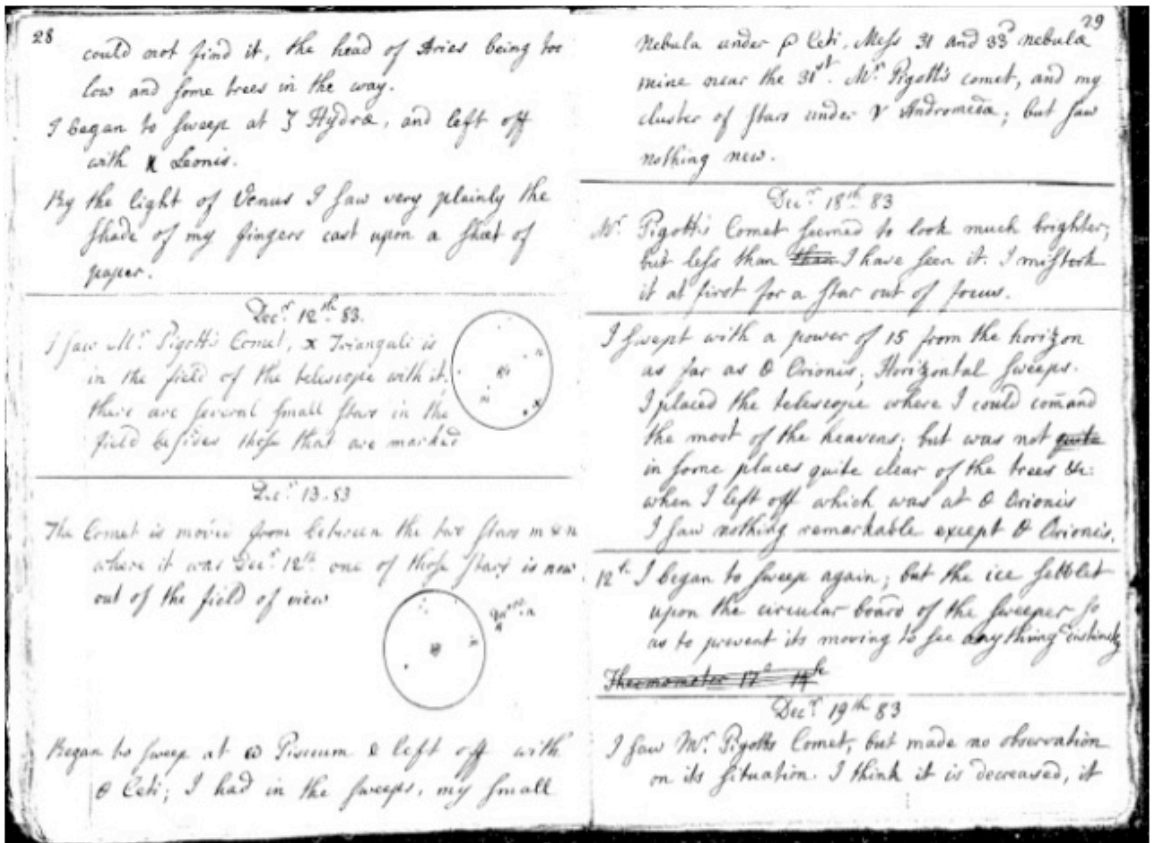

Figure 5.   Caroline Herschel, Piggott's Comet, MS. RAS C.1/1.28, 1783, Royal Astronomical Society, London

### 3. Caroline and the Comet-Crazy Centuries

Caroline Herschel made her astronomical contributions during an exciting era in post-Enlightenment Georgian England when, despite the predicted return of Halley's Comet in 1759, comets were still being separated from other transient denizens of the solar system, like meteors. Comet fever was rampant, and the heavens were figuratively on fire. Extraordinary progress in the field of optics and physics aided in the sighting of at least 62 comets during the eighteenth century, whose last quarter featured one comet annually. Since science was a form of serious entertainment, devices called 'cometariums' allowed popularizing astronomers to showcase the movement of comets and theories about them in spectacles all over Great Britain.[xii] Because of their topical natures, comets were also frequently employed as satirical devices during a era that witnessed the ascendancy of journalism and caricature. Caricaturists mined these rhetorical devices for their rich satirical potential (Figure 6). Caroline herself helped create the frenzy surrounding comets. She caught the attention of the media, as it existed then, not like Lady Gaga of today but still grist for the mill of satire (Figure 7).[xiii] This print, which lampoons the confusion between comets and meteors as well as female astronomers, appeared a month after Caroline had discovered her third comet.

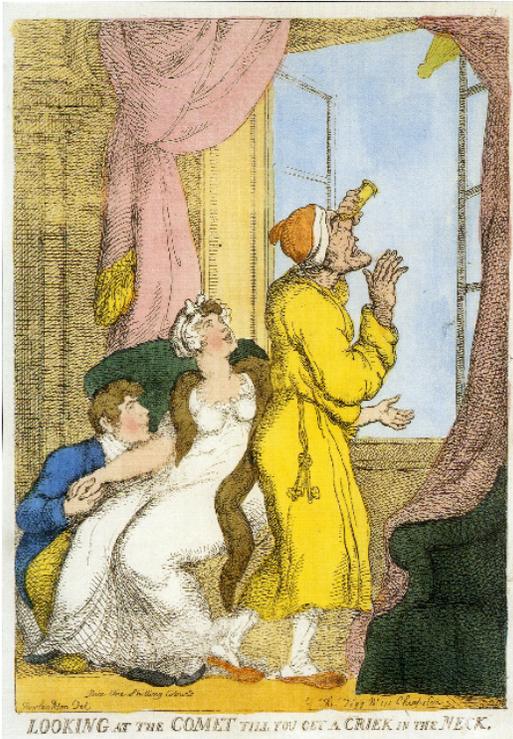

Figure 6.    Thomas Rowlandson, Looking at the Comet Till You Get a Criek in the Neck, 1811, hand-coloured etching, Private Collection

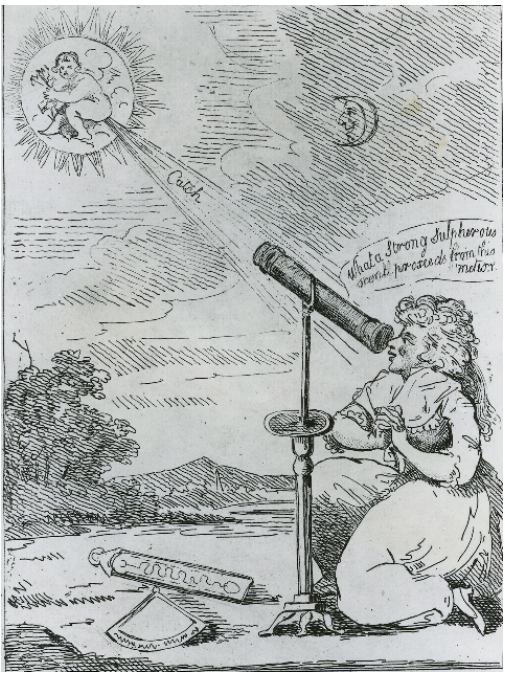

Figure 7.    R. Hawkins ? (in the style of Rowlandson), *The Female Philosopher Smelling out the Comet*, 1790, etching, Morgan Library and Museum, New York, Peel Collection Vol. IX

**4. Caroline's Comets**

From Slough, industrious Caroline discovered eight comets—five of which were published in the *Philosophical Transactions*. Her correspondence about each is contained in separate files in the Royal Astronomical Society.[xiv]

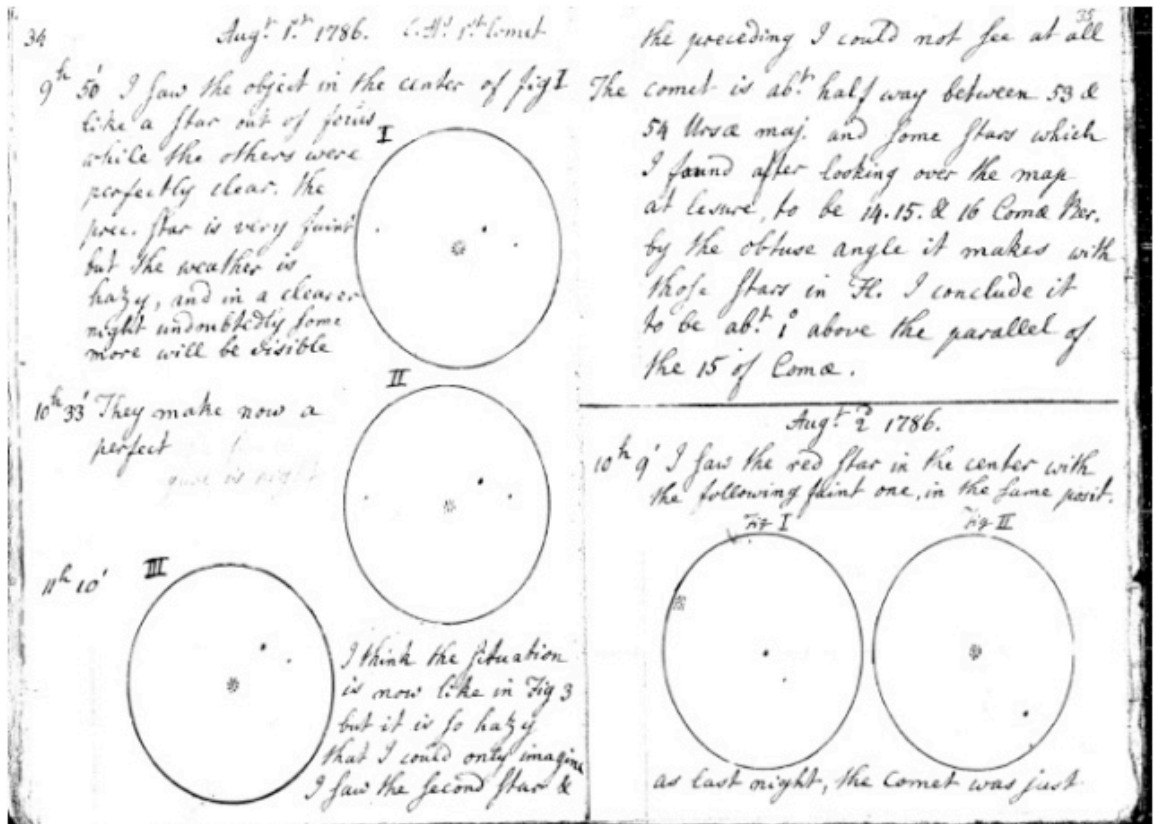

Figure 8. Caroline Herschel, Comet C/1786 P1 (Herschel), MS. RAS C.1/1.1, 34-35, 1786, Royal Astronomical Society, London

Caroline discovered her first comet on 1 August 1786, Comet C/1786 P1 (Herschel), while William was in Göttingen delivering a telescope (Figure 8). She found it with her small sweeper, documenting it in her first record book from 1 August to 12 November.[xv] Unable to calculate the mathematical coordinates of this celestial object, she accompanied her observations of the first night with three drawings of its position in viewing fields. On the following evening, she drew it twice and confirmed its cometary nature: '1 o'clock; the object of last night *is a Comet*. I did not go to rest till I had wrote to Dr. Blagden and Mr. Aubert to announce the comet'. She stated, 'In consequence of the friendship which I know to exist between you and my Brother, I venture to trouble you in his absence with the following imperfect account of a comet. . . .' Aubert replied: 'You have immortalized your name. . . .' On 6 August Caroline entertained the president and secretary of the Royal Society and members of the London intelligentsia, who had journeyed to Slough for the sole purpose of seeing Caroline's comet through Caroline's telescope! William returned to discover that his sister was famous, and was summoned to Windsor Castle to demonstrate Caroline's comet to the royal family. William also recorded the phenomenon, terming it 'My Sister's Comet'.[xvi] Caroline's letter to Blagden was published in the *Philosophical Transactions*—together with six figures based on Caroline's drawings but more systematically plotted—in which she noted that she used a 'Newtonian sweeper of 27 inches focal length, and a power of about 20'.[xvii] William contributed an addendum and his own remarks.[xviii] Previously, on 19 November William had Caroline annotate on a separate paper attached to folio 45 of her record book facts about the comet and a diagram that became the sixth figure. Inserted after folio 46 are two additional diagrams with the comet's position on 31 August and 1 September. Francis Wollaston also published his observations and calculations of 'Miss Herschel's Comet' in this volume.[xix]

　　With the courtship, beginning in 1786, and the marriage on 8 May 1788 of 50-year-old William to their neighbor, the wealthy widow Mary Burney Pitt, his enthusiasm for observing waned, and Caroline had more time to use her sweeper. Feeling displaced, she demanded that her brother ask the King to pay her salary for her work. Apparently, Joseph Banks, President of the Royal Society, advised that it should come officially from Queen Charlotte, as Caroline was 'the ladies' comet hunter'. William reasoned, in a letter of August 1787 to Banks, that his sister was the best possible assistant, and if she were to decline, he would have to spend £100 more for another assistant.[xx] Caroline was granted an annual stipend of £50 (female servants earned about £10 per annum, governesses around £40), the first professional salary ever paid a woman scientist in Britain, marking a social revolution.[xxi] Caroline claimed it was 'the first money I ever in all my lifetime thought myself to be at liberty to spend to my own liking'.[xxii] This money gave her a sense of independence. In an age when no female was expected to engage in scientific research, she was an internationally acclaimed curiosity. She remains an early role model of what can be achieved with limited opportunities and selective dedication.

　　Caroline's next discovery came on 21 December 1788, Comet 35P/1788 Y1 (Herschel-Rigollet), which she observed until 8 January 1789.[xxiii] As with each of her comets, she annotated it with 'C.H.' and its number (Figure 9). On the subsequent page, she noted a mistake in her sweeping in search of it, adding that Nevil Maskelyne, the Astronomer Royal—who was fast becoming her good friend and later would address her as 'Dear Sister Astronomer'—had last observed it on 4 February. (The RAS preserves a letter of 22 December from Caroline to Maskelyne announcing its discovery with a note by William.) William, who had tracked it with his ten-foot reflector, wrote a letter about it published in the *Philosophical Transactions*, in which he reported that Caroline used her small Newtonian sweeper.[xxiv] In 1939, after Roger Rigollet in France rediscovered it, this comet was deemed to be periodic and named Comet Herschel-Rigollet.

　　Caroline was not always right about her comet identifications, as in her 18 August 1789 claim—which she corrected the following night.[xxv] Also, in 1794, she reported a comet that turned out to be a light in a barn, according to Wollaston![xxvi]

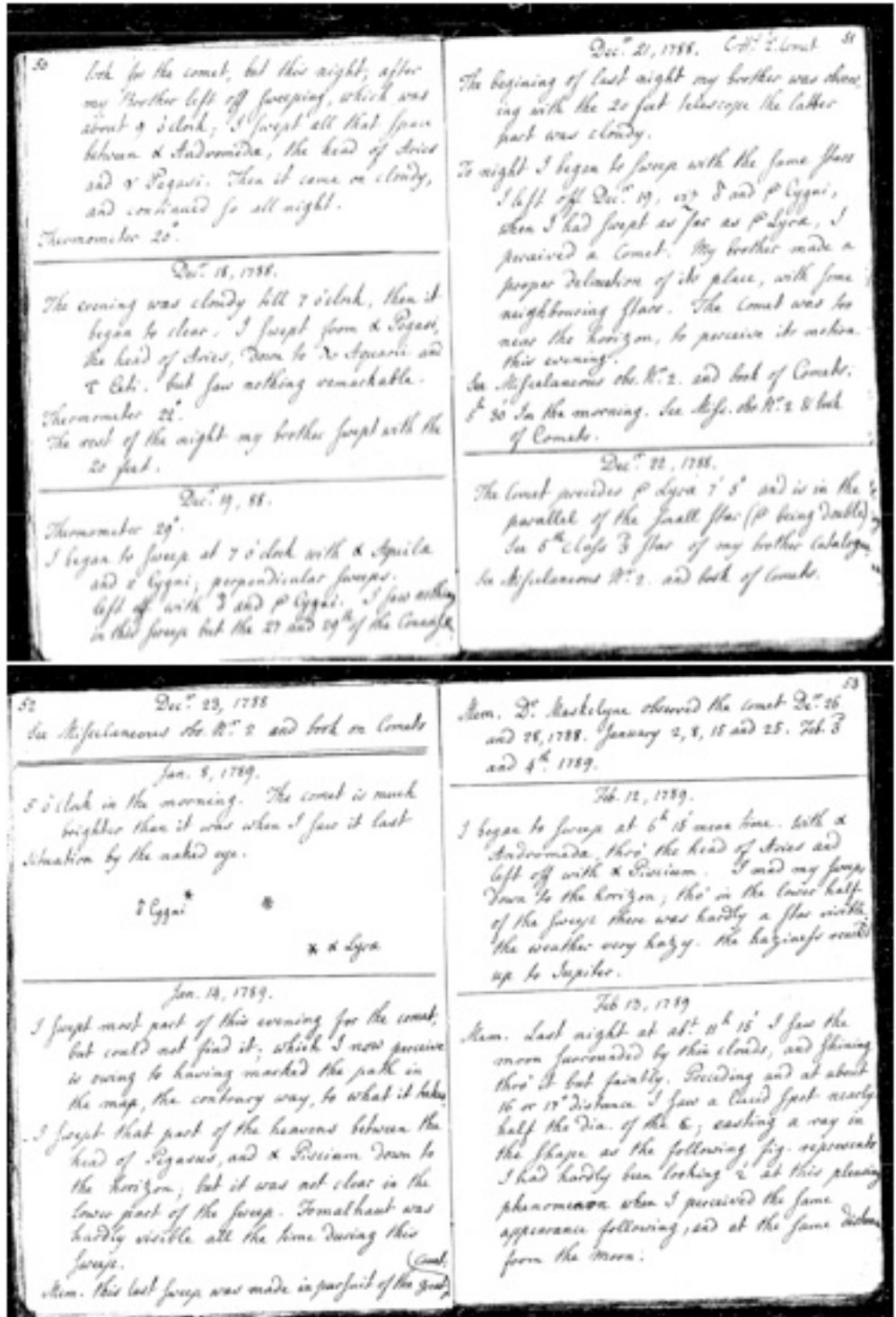

Figure 9.   Caroline Herschel, Comet 35P/1788 Y1 (Herschel-Rigollet), MS. RAS C.1/1.2, 50-54, 1788, Royal Astronomical Society, London

Her third comet discovery, Comet C/1790 A1 (Herschel), came on 7 January 1790 (Figure 10). However, the apparition was poor, and Caroline never sighted it again due to cloudy skies. On 19 January, she wrote not without some frustration: 'I have swept all evening for my comet in vain'.[xxvii] William also noted the 'supposed Comet'.[xxviii]

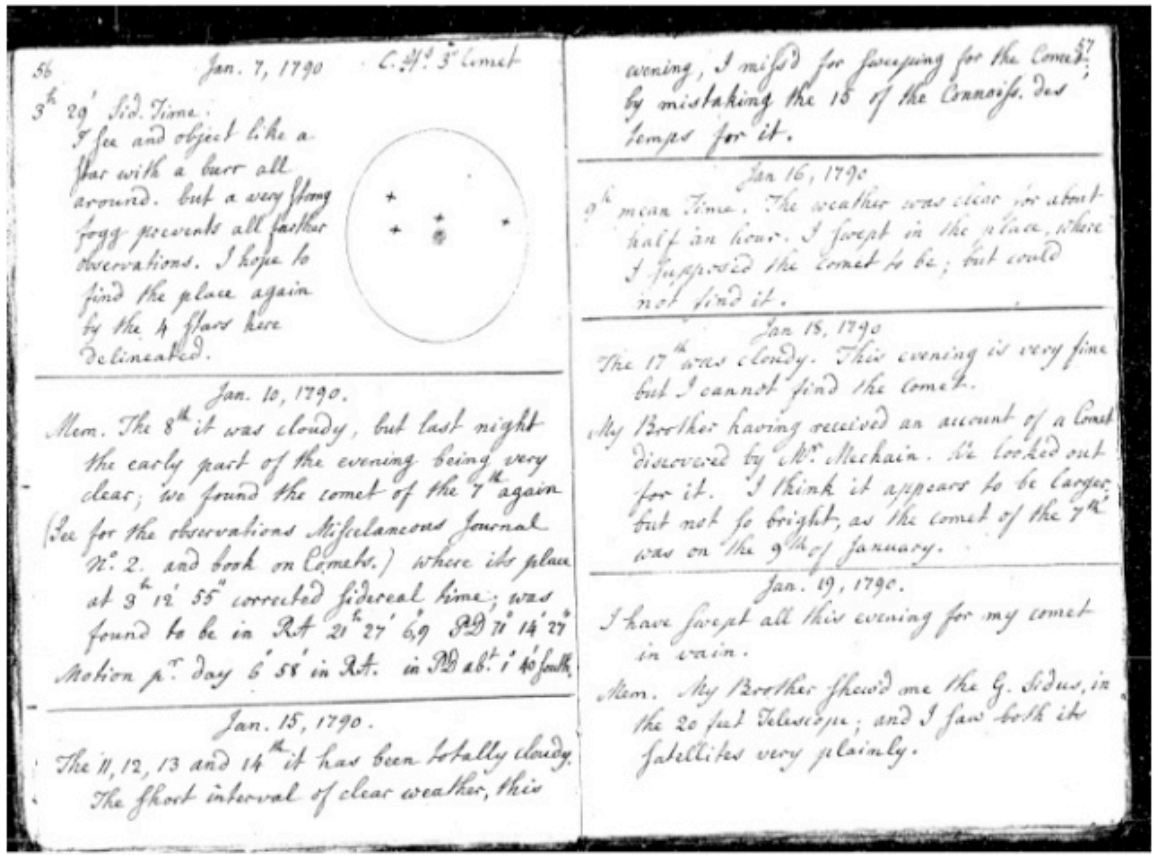

Figure 10.   Caroline Herschel, Comet C/1790 A1 (Herschel), MS. RAS C.1/1.2, 56, 1790, Royal Astronomical Society, London

1790 was a good year for Caroline. Her fourth discovery, Comet C/1790 H1 (Herschel), her second for that year (Figure 11), occurred on 17 April (not the 18th as frequently recorded) and was observed by her until 10 June.[xxix] William confirmed it.[xxx] By the beginning of May, Comet Herschel developed a visible tail. Previously, Maskelyne had been disappointed with her reporting and calculations. But, using the measuring device on her new five-foot sweeper—the triangle improved her observations—and clearly excited by them, she could sweep vertically, as well as horizontally, on the horizon at sunset or sunrise, making it ideal for comets. Many evenings she did not record which sweeper she used but most probably employed the smaller for sweeping and the larger for calculating exact positions as she could only use it standing on a stool to focus. The subsequent pages of her observation book are filled with complicated calculations demonstrating her development—heating up around perihelion, so to speak.

On 15 December 1791, breaking a 20-month drought of comet discoveries by anyone, Caroline spotted her fifth comet, Comet C/1791 X1 (Herschel). Although she did not draw it, she commented about it, noting that her brother had spied it the same night with the 'seven feet reflector'.[xxxi] William confirmed it and published the discovery, crediting it to his sister with her 'five-feet Newtonian Sweeper', in the *Philosophical Transactions*.[xxxii]

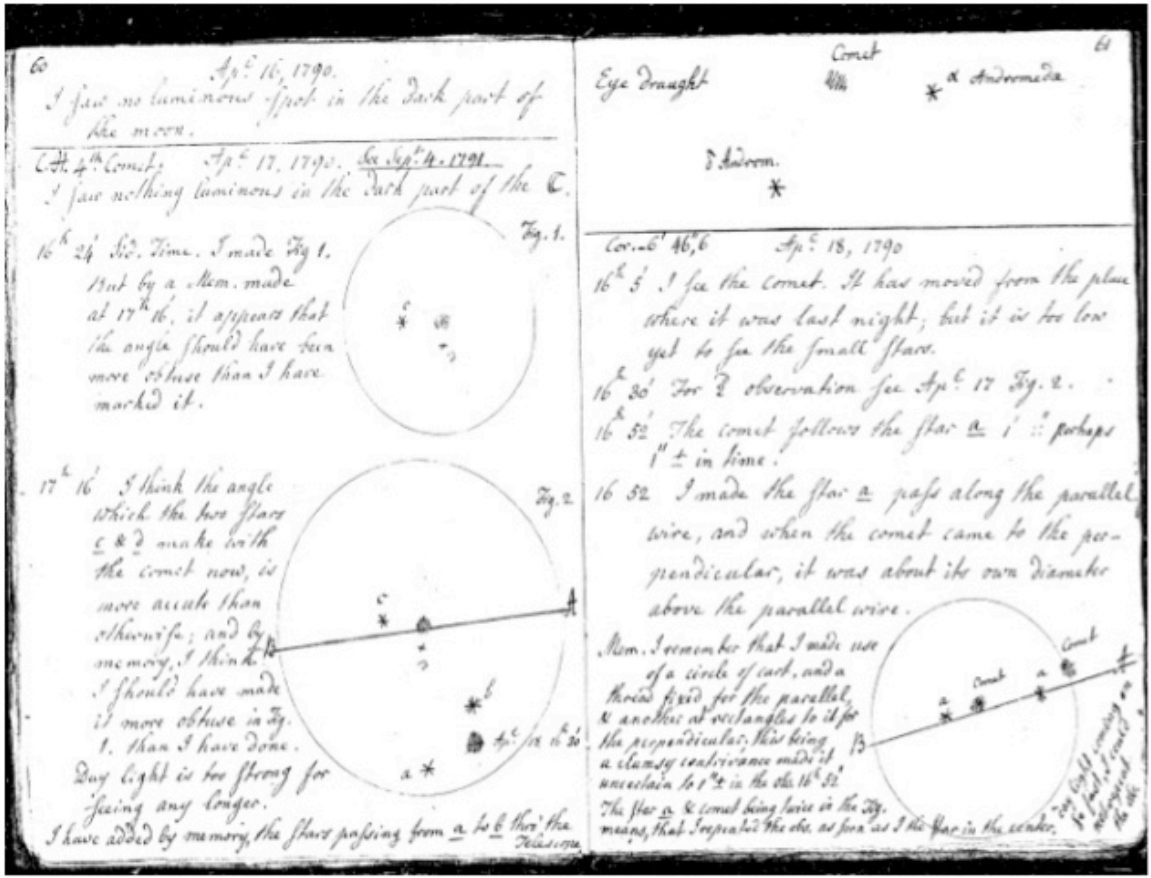

Figure 11.   Caroline Herschel, Comet C/1790 H1 (Herschel), MS. RAS C.1/1.2, 60-61, 1790, Royal Astronomical Society, London

Caroline discovered her sixth comet on 7 October 1793 but did not draw it, although William confirmed it. Unknown to her, Messier had sighted it earlier, and thus it was named after him: Comet C/1793 S2 (Messier). Nonetheless, she later annotated her entry for that night 'C.H. 6th Comet'. Caroline officially reported on it in a letter of 8 October, which was subsequently published in the *Philosophical Transactions*.[xxxiii]

On 7 November 1795, Caroline 'discovered' a comet that she labeled 'C.H. 7th Comet' (Figure 12), and her brother confirmed 'my Sister's Comet' that evening, noting that it was visible to the naked eye.[xxxiv] Nearly ten years before, on the evening of 17 January 1786, Caroline had spotted this same comet, which had been first sighted by the famous French comet hunter Pierre Méchain. Since it had been observed for only two subsequent nights, no orbit could be calculated. Not associating the two apparitions, she announced the comet discovery in a letter, to which William appended additional observations with his calculations, which was published in the *Philosophical*

*Transactions*, wherein she noted that she used 'my 5-feet reflector' in her sweeping.[xxxv] Ten years later, Johann Encke suggested incorrectly that this comet had an elliptical orbit with a period of 12.1 years, although its next apparition in 1822 was accurately predicted. It was only the second comet for which a return had been successfully predicted, the first being Halley, and thus the comet was named after Encke (Comet 2P/Encke.)[xxxvi] However, this periodic comet could have easily had an alternate name that included Herschel.

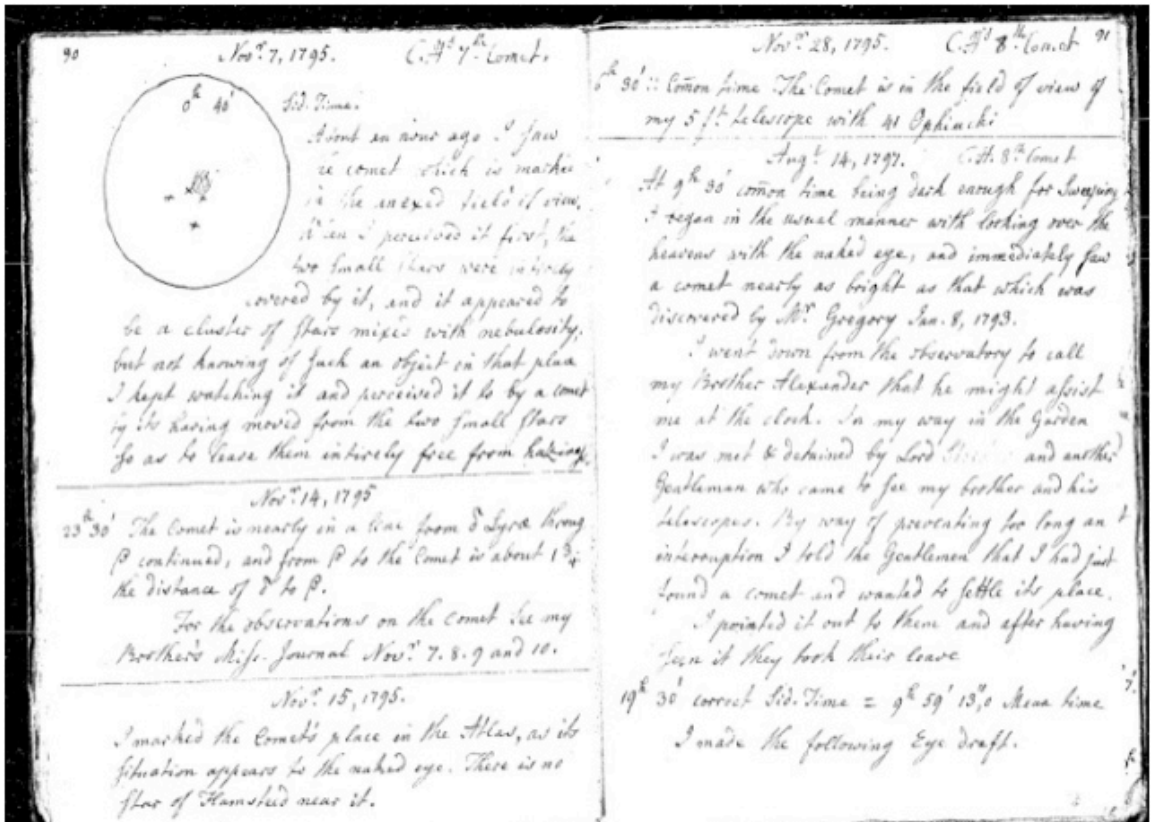

Figure 12.   Caroline Herschel, Comet 2P/Encke, MS. RAS C.1/1.2, 90, 1795, Royal Astronomical Society, London

Caroline's eighth and generally accepted final comet discovery (Figure 13) occurred on 14 August 1797 with the naked eye, and she observed it through 23 August, the first day William noted it as well. She and Alexis Bouvard in Paris independently discovered Comet C/1797 P1 (Bouvard-Herschel).[xxxvii] To avoid the delay of a messenger or the post, as with her second comet, she slept one hour, saddled a horse, and rode from Slough to Greenwich to notify Maskelyne. This six-hour, over 26 mile trip, was grueling, let alone dangerous at night in a sidesaddle, especially since she was accustomed to riding at the maximum for two hours. To recover, she stayed several days with the Maskelynes and confessed all in an apology to Banks.

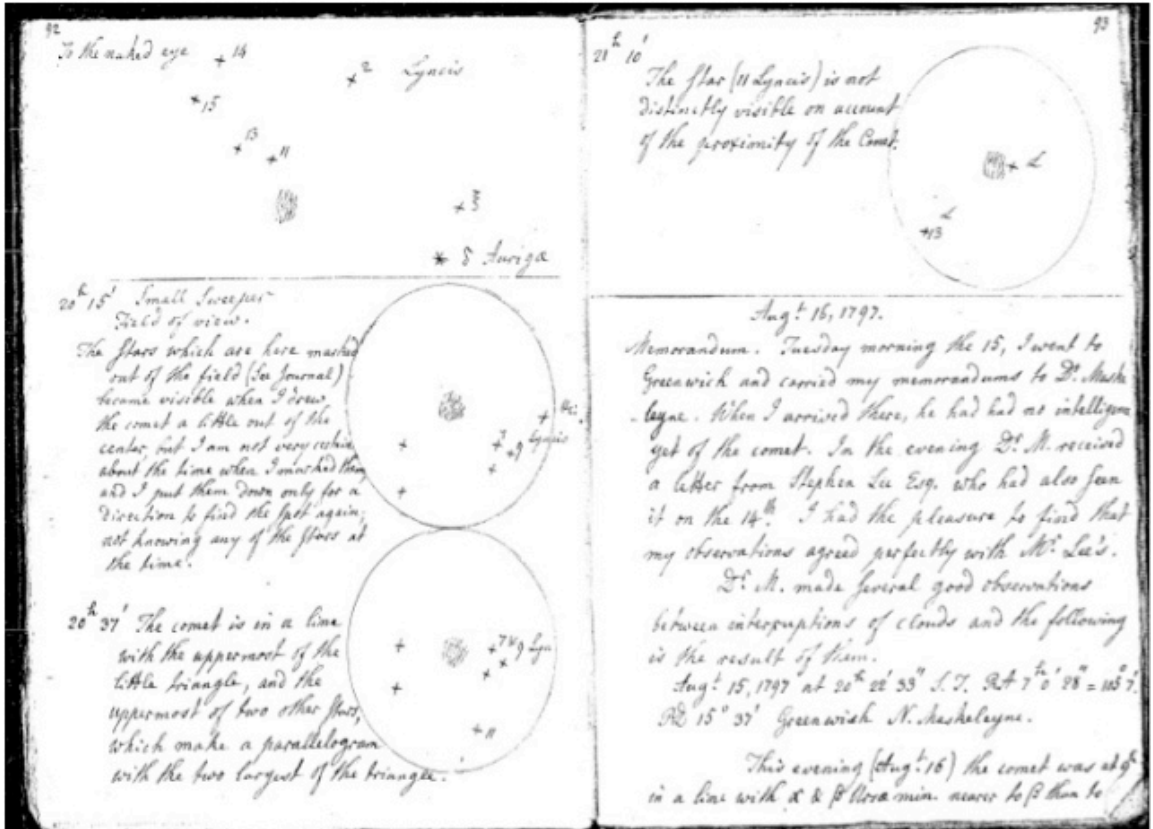

Figure 13.    Caroline Herschel, Comet C/1797 P1 (Bouvard-Herschel), MS. RAS C.1/1.2, 92-93, 1795, Royal Astronomical Society, London

Soon thereafter, as an assertion of professional independence and perhaps even in a recognition of a certain rivalry with her brother, she made the disastrous decision to leave the cottage at Observatory House and live in a series of rental lodgings near Windsor. She left a soulless mention that does not reveal what complex motives provoked her, which no doubt were detailed in diaries that she destroyed. With her pension she could be more independent, especially as William, less interested in astronomy, by 1799 had turned to geology and was spending more time in Bath. She may have hoped that living in lodgings would have ameliorated her loneliness. Concurrently, William referred to her more regularly in his papers as 'my indefatigable assistant Caroline Herschel' or 'my sister, Miss Herschel'. They continued working together but instead of shouting, they devised a system of coded rope pulls, hand signals, and bells, later adding a flexible speaking tube.

While Caroline recorded other comets, as did William, she did not draw any others until 1799 and 1807.[xxxviii] After reading about it, she looked for, found, and sketched the latter, C/1806 VI (Pons).[xxxix] Later that year, she sketched the Great Comet of 1807 (C/1807 R1); her drawing from 24 October 1807 (Figure 14) demonstrates how much Caroline had developed her observational skills as well as her draftsmanship. This more detailed rendering is atmospheric, illusionistic, and three-dimensional—capturing the vaporous nature of the comet's tail brilliantly. One wonders where she had acquired this ability? Did she study the drawings of other astronomers, which likewise were becoming more accurate, or through contact with artists? Arguably, the comet's size and its lengthy apparition facilitated detailed study. Caroline observed the comet from 2 October through 24 October 1807, while William recorded it from 4 October 1807 until 21 Feb. 1808.[xl]

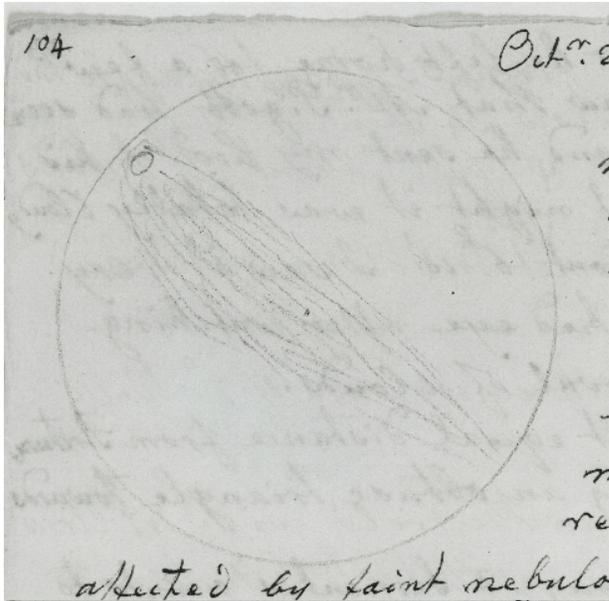

Figure 14.    Caroline Herschel, The Great Comet of 1807 (C/1807 R1), MS. RAS C1/1.3, 104 (detail), 1807, Royal Astronomical Society, London

Until the appearance of C/1995 O1 (Hale-Bopp), the Great Comet of 1811 (C/1811 F1 [Flaugergues]) held the record as the longest naked-eye comet visible—over a year telescopically and nearly a year naked eye. No wonder that both Herschels—Caroline in Figure 15 and William in Figure 16—made detailed portraits of it. Caroline observed it from 1 September through 18 September 1811, writing only a few notes about it, while William recorded it from 2 September until 23 December 1811.[xli] As it had the largest coma ever recorded, as well as a very long tail, both Herschels delineated its coma! The siblings drew the comet's entire anatomy, including its center of condensation and even its nucleus. Clearly, these sketches are watershed works, reflecting William's ideas that comets have atmospheres of nebulous material around their heads, which subsequently has been proven to be essentially correct. William published major reports on both the comets of 1807 and 1811. But Caroline's style is even more advanced than William's, because she used graphite to suggest the comet's diaphanous coma and tail, and pen and ink to outline its head and nucleus, making one wish for a fuller explanation of her great leap.

Artists of the period were mesmerized by comets and, like astronomers, had acute visual skills. For example, the painter John Linnell was fascinated by the Great Comet of 1811 and plotted its course in his journal and in several drawings, two reproduced here (Figures 17-18).[xlii] With his mastery of media, he used white chalk to best capture the comet's diaphanous appearance in these two sheets preserving its nocturnal magic.

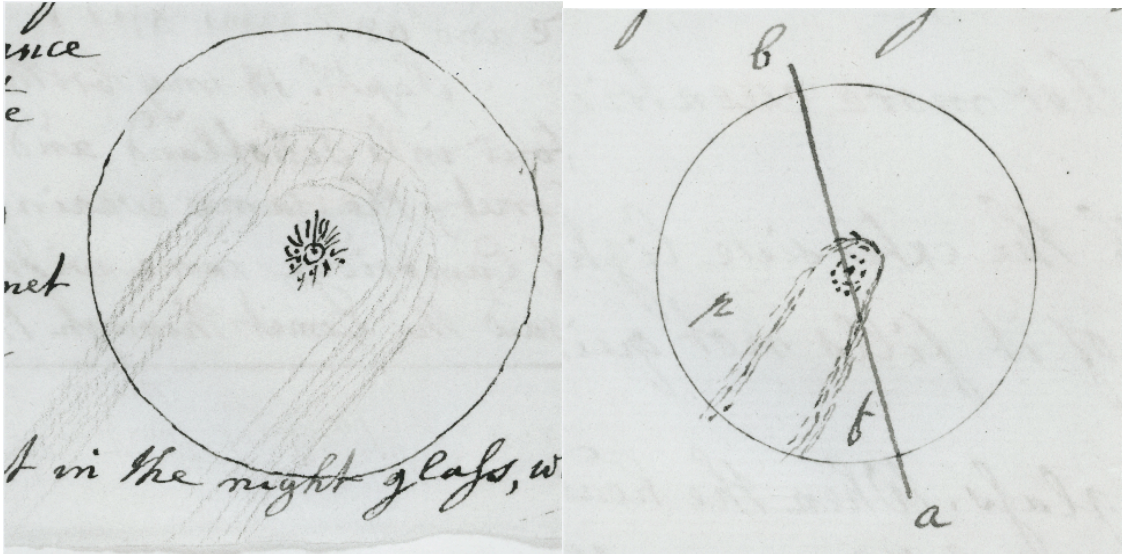

Figure 15.  (left)    Caroline Herschel, The Great Comet of 1811 (C/1811 F1), MS. RAS C.1/1.3, 105 (detail), 1811, Royal Astronomical Society, London

Figure 16. (right)   William Herschel, The Great Comet of 1811 (C/1811 F1), MS. RAS W.3/1.12, 37 (detail), 1811, Royal Astronomical Society, London

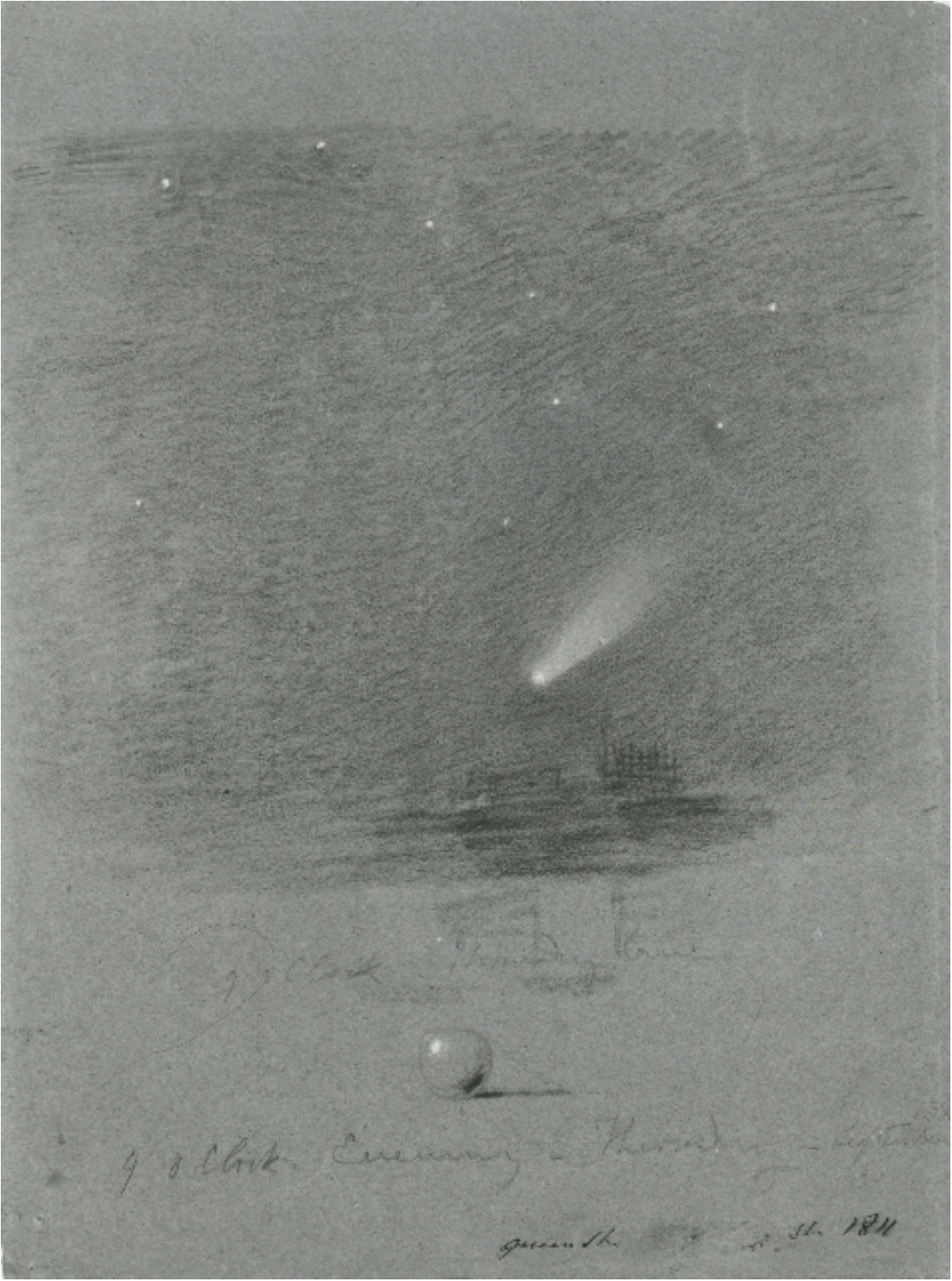

Figure 17. John Linnell, Two Studies of the Great Comet of 1811 (C/1811 F1), 1811, white chalk on gray paper, The British Museum, London

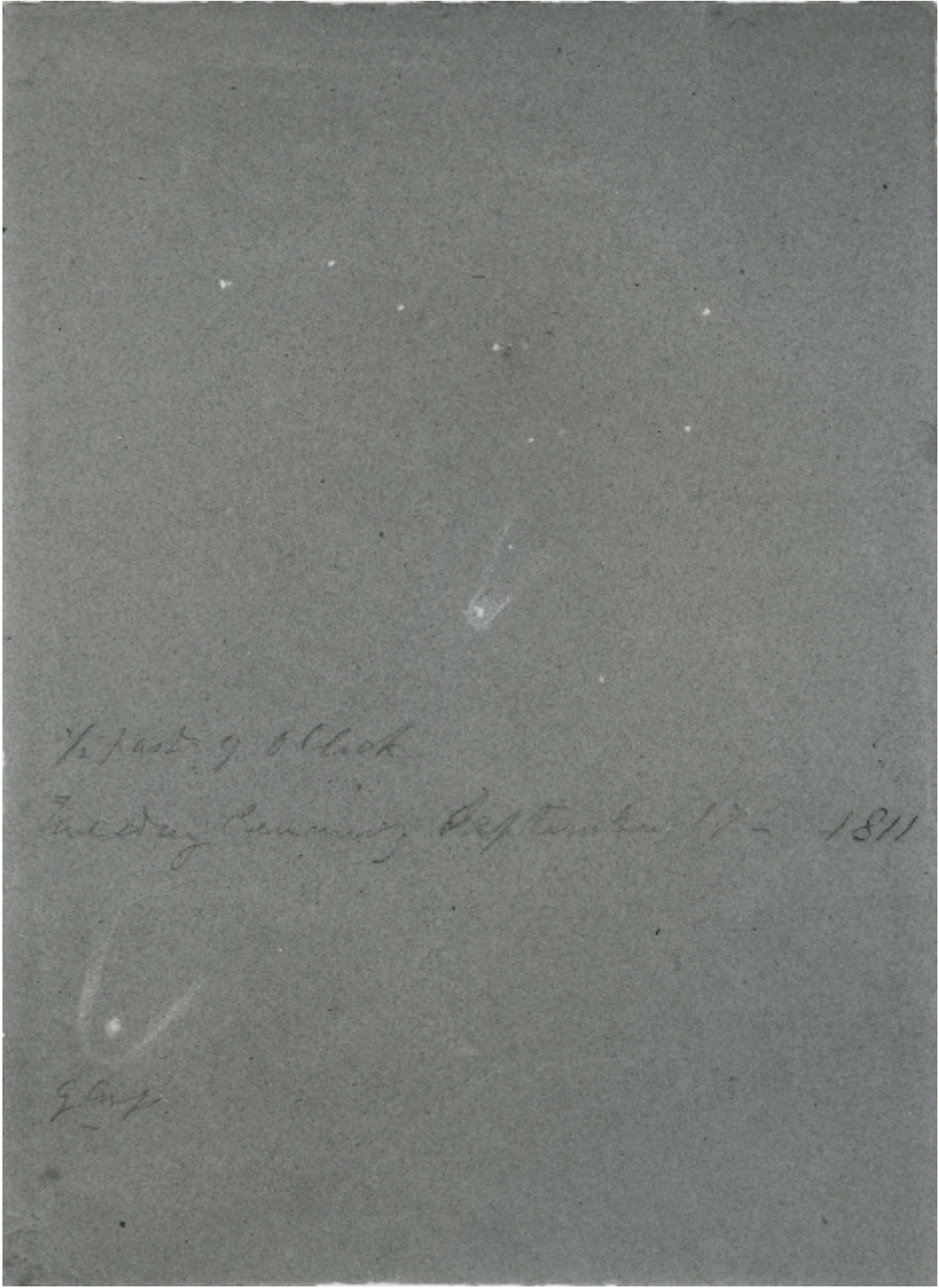

Figure 18.   John Linnell, The Great Comet of 1811 (C/1811 F1), 1811, black and white chalk on gray paper, The British Museum, London

After 1811, Caroline continued to sweep for comets, but she had stiff competition. Interestingly, a perusal of her manuscripts reveals no interest in comets beyond the mere act of discovery—no real astronomical thoughts about their physical nature, their past or future apparitions, or philosophical speculations. Once she had found a comet, she handed it over to the experts: her brother William and Maskelyne. However, her qualities of dedication, attention to detail, accuracy, and perseverance, equipped her for the labor-intensive sweeping of the skies that had few sensational rewards.

Following William's death in 1822, Caroline went back to her family in Germany, an impetuous decision she later regretted. At William's death, her nephew, John Hershel, a prodigy and polymath, took over observing at Slough. Caroline had given John his first introduction in astronomy, showing him the constellations in John Flamsteed's *Atlas*. The torch had been passed, and, among his many other achievements, John further developed the portrayal of comets during a time of sophisticated pre-photographic reproductive techniques. In Hanover, Caroline closed her observing book on 31 January 1824 with a sad, perfunctory entry about the Great Comet of 1823, C/1823 Y1, which she spied from her lodgings ('observed here at Hannover') with her small sweeper (Figure 19). It had already been discovered on 29 December 1823.[xliii]

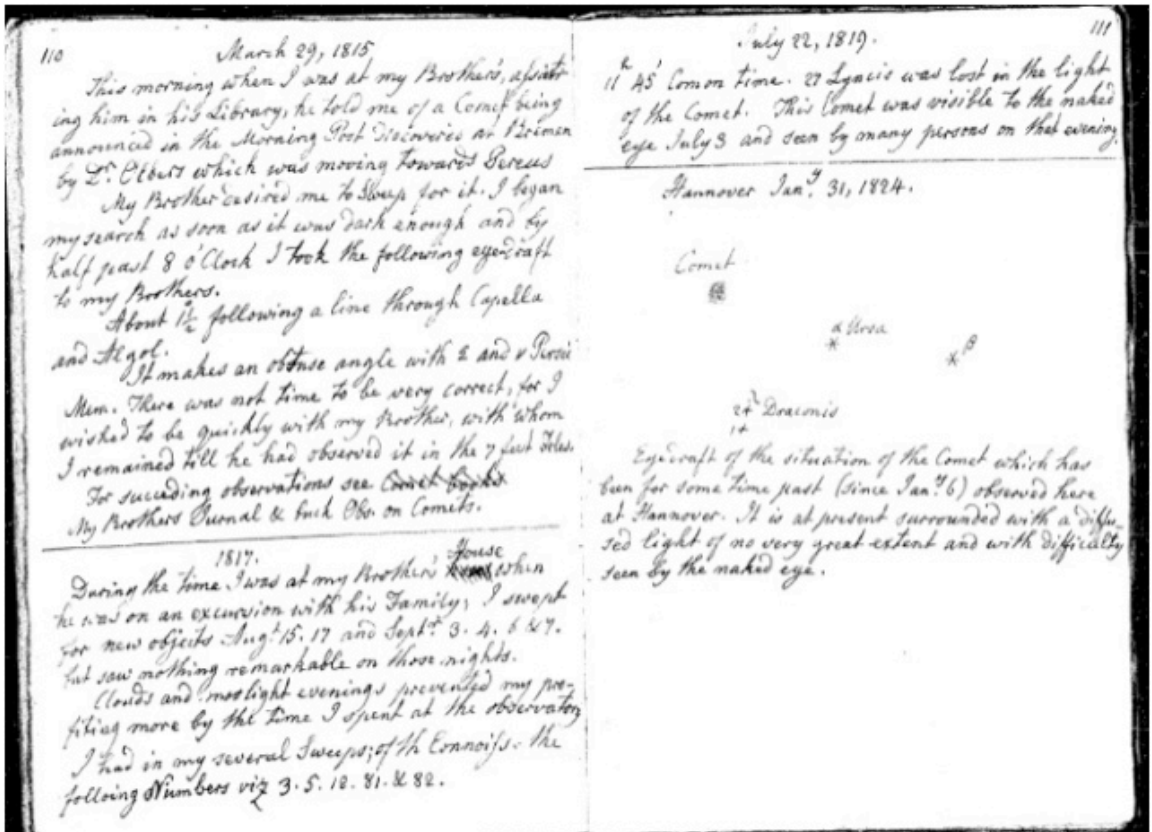

Figure 19.    Caroline Herschel, the Great Comet of 1823 (C/1823 Y1), MS. RAS C.1/1.3, 111, 1824, Royal Astronomical Society, London

In Hanover, Caroline lived out her life as a local celebrity, which amazed and amused her, continuing to garner accolades. In 1828, the Royal Astronomical Society awarded her their Gold Medal for her work reformulating the 2500 nebulae previously catalogued by William and her into a format that John could use for his nightly observations revising it; in 1832, the King of Denmark honored her earlier comet discoveries with a medal. These accolades were followed in 1835 by an honorary membership from the Royal Astronomical Society, which they also bestowed on Mary Somerville, a popularizer of astronomy, making these two the first of their sex to ever to receive this honor. In 1838, the Royal Irish Academy followed suit, and in 1846, the King of Prussia bestowed on Caroline the gold medal for science. Her other honors include Asteroid 281, discovered in 1888 and named Lucretia, and the crater C. Herschel on the moon. All things considered, Caroline's career was rather like that of a comet. At first her glory reflected that of her brother—as a comet reflects the light of the sun—but as she approached perihelion, she burned brightly not only with William's light but also with the light of her own achievements. Further, like a comet in its orbit around the sun, her career blazed briefly—for only about 10 years.

Caroline's comet discoveries created her legacy, establishing a precedent for female astronomers, and her record for comet discoveries remained unequaled for female astronomers until Carolyn Shoemaker in the 1980s. Caroline Herschel worked as an astronomer at a time when discovering comets and calculating their orbits was an obsession, not only for astronomers but also the public. She became the Diana of the comet world, the huntress of comets, holding the record for comet discoveries by a woman—eight—until April 1987, when Shoemaker surpassed her record on photographic plates rather than with the eye, though Carolyn Shoemaker has also been slighted by history in the Shoemaker-Levy comets (number 9 famously crashed into Jupiter), which should be recorded as C. Shoemaker-E. Shoemaker-Levy. And, perhaps there was even a ninth comet that Caroline Herschel identified only as a 'rich spot' on 30 July and 24 August 1783.[xliv] In any case, the record for live observations really remains in Caroline Herschel's court.

**Figures**

Figure 1.   Silhouette of Caroline Herschel, c. 1768, MS. Gunther 36, fol. 146r © By permission of the Oxford University Museum of the History of Science

Figure 2.   Paul Sandby, The Meteor of 18 August 1783, 1783, aquatint, The British Museum, London

Figure 3.   Thomas or Paul Sandby, The Meteor 18 August 1783, 1783, watercolor over graphite, The British Museum, London

Figure 4.   Caroline's Sweeper, c. 1783, Historisches Museum, Hanover

Figure 5.   Caroline Herschel, Piggott's Comet, MS. RAS C.1/1.28, 1783, Royal Astronomical Society, London

Figure 6.    Thomas Rowlandson, Looking at the Comet Till You Get a Criek in the Neck, 1811, hand-coloured etching, Private Collection

Figure 7.    R. Hawkins ? (in the style of Rowlandson), *The Female Philosopher Smelling out the Comet*, 1790, etching, Morgan Library and Museum, New York, Peel Collection Vol. IX

Figure 8.    Caroline Herschel, Comet C/1786 P1 (Herschel), MS. RAS C.1/1.1, 34-35, 1786, Royal Astronomical Society, London

Figure 9.    Caroline Herschel, Comet 35P/1788 Y1 (Herschel-Rigollet), MS. RAS C.1/1.2, 50-54, 1788, Royal Astronomical Society, London

Figure 10.   Caroline Herschel, Comet C/1790 A1 (Herschel), MS. RAS C.1/1.2, 56, 1790, Royal Astronomical Society, London

Figure 11.   Caroline Herschel, Comet C/1790 H1 (Herschel), MS. RAS C.1/1.2, 60-61, 1790, Royal Astronomical Society, London

Figure 12.    Caroline Herschel, Comet 2P/Encke, MS. RAS C.1/1.2, 90, 1795, Royal Astronomical Society, London

Figure 13.    Caroline Herschel, Comet C/1797 P1 (Bouvard-Herschel), MS. RAS C.1/1.2, 92-93, 1795, Royal Astronomical Society, London

Figure 14.    Caroline Herschel, The Great Comet of 1807 (C/1807 R1), MS. RAS C1/1.3, 104 (detail), 1807, Royal Astronomical Society, London

Figure 15.    Caroline Herschel, The Great Comet of 1811 (C/1811 F1), MS. RAS C.1/1.3, 105 (detail), 1811, Royal Astronomical Society, London

Figure 16.    William Herschel, The Great Comet of 1811 (C/1811 F1), MS. RAS W.3/1.12, 37 (detail), 1811, Royal Astronomical Society, London

Figure 17.     John Linnell, Two Studies of the Great Comet of 1811 (C/1811 F1), 1811, white chalk on gray paper, The British Museum, London

Figure 18.     John Linnell, The Great Comet of 1811 (C/1811 F1), 1811, black and white chalk on gray paper, The British Museum, London

Figure 19.     Caroline Herschel, the Great Comet of 1823 (C/1823 Y1), MS. RAS C.1/1.3, 111, 1824, Royal Astronomical Society, London

**Acknowledgments**. We would like to thank the late Peter Hingley, Librarian, Royal Astronomical Society; Michael Hoskin, Churchill College, Cambridge; and Madeline Kennedy, Williams College.

---

[i] Among his many publications on the Herschels that we consulted are: Michael Hoskin, *The Herschel Partnership: As Viewed by Caroline* (Cambridge: Science History Publications, 2003); Michael Hoskin, 'Caroline Herschel as Observer', *Journal for the History of Astronomy*, vol. 35 (2005), pp. 373-405; Michael Hoskin, *Discoverers of the Universe: William and Caroline Herschel* (Princeton: Princeton University Press, 2011); Michael Hoskin (ed.), *Caroline Herschel's Autobiographies* (Cambridge: Science History Publications, 2003).

[ii] Roberta J. M. Olson and Jay M. Pasachoff, *Fire in the Sky: Comets and Meteors, the Decisive Centuries, in British Art and Science* (Cambridge: Cambridge University Press, 1998).

[iii] Caroline Herschel, *Caroline Herschel's Autobiographies*, ed. Michael Hoskin, (Cambridge: Science History Publications, 2003), p. 24.

---

[iv] See Roberta J. M. Olson and Jay M. Pasachoff, 'The "wonderful meteor" of 18 August 1783, the Sandbys, "Samuel Scott", and heavenly bodies', *Apollo*, vol. 146, no. 429 (1997), pp. 12-19; Olson and Pasachoff, *Fire in the Sky*, pp. 63-78.

[v] Johnson Ball, *Paul and Thomas Sandby, Royal Academicians* (Somerset, Charles Skilton Ltd., 1985), pp. 85, 246-47. In 1789, Thomas Sandby and his family were invited to the dinner inside the forty-foot telescope at Slough.

[vi] Michael Hoskin, 'Caroline Herschel's "Small" Sweeper', *Journal for the History of Astronomy*, vol. 36 (2005), pp. 28-30.

[vii] Herschel, William, 'Catalogue of One Thousand New Nebulae and Clusters of Stars', *Philosophical Transactions of the Royal Society*, vol. 76 (1786), pp. 457-99; William Herschel, 'Catalogue of a Second Thousand of New Nebulae and Clusters of Stars; with a Few Introductory Remarks on the Construction of the Heavens', *Philosophical Transactions of the Royal Society*, vol. 79 (1789), pp. 212-55; William Herschel, 'Catalogue of 500 New Nebulae, Nebulous Stars, Planetary Nebulae, and Clusters of Stars; with Remarks on the Construction of the Heavens', *Philososphical Transactions of the Royal Society*, 92 (1802), pp. 477-528.

[viii] Robert Holmes, *The Age of Wonder: How the Romantic Generation Discovered the Beauty and Terror of Science* (New York: Vintage Books. A Division of Random House, Inc., 2008), p. 117.

[ix] MS. RAS C.1 (wrapper), C.1.1.1, Royal Astronomical Society, London (hereafter cited only as RAS).

[x] MS. RAS C.1/1.1, 27-30.

[xi] MS. RAS W.3/1.12, 1-2.

[xii] See Olson and Pasachoff, *Fire in the Sky*, pp. 44-48.

[xiii] For a more in-depth consideration of caricature and the first publication of the satire of Caroline, see *ibid*., pp. 132-60, fig. 72.

[xiv] MS. RAS C.1/3,1-8.

[xv] MS. RAS C.1/1.1, 34-45; G. W. Kronk, *Cometography*: *A Catalog of Comets*, vol. 1 (Cambridge: Cambridge University Press, 2000), pp. 482-84.

[xvi] MS. RAS W.3/1.12, 3.

[xvii] Caroline Herschel, 'An Account of a New Comet', *Philosophical Transactions of the Royal Society*, vol. 77 (1787), pp. 1-3.

---

[xviii] William Herschel, 'Remarks on A New Comet', *Philosophical Transactions of the Royal Society*, vol. 77 (1787), pp. 4-5.

[xix] Francis Wollaston, 'Observations of Miss Herschel's Comet, in August and September', *Philosophical Transactions of the Royal Society*, vol. 77 (1787), pp. 55-60.

[xx] MS. RAS W.1/5.1v, 2.

[xxi] Holmes, *The Age of Wonder*, p. 179.

[xxii] C. Herschel, *Autobiographies*, p. 94.

[xxiii] MS. RAS C.1/1.2, 51-52; Kronk, *Cometography*, vol. 1, pp. 487-89.

[xxiv] MS. RAS W.3/1.12, 4-6; Herschel, William, 'Observations on a Comet', *Philosophical Transactions of the Royal Society*, vol. 79 (1789), pp. 151-53.

[xxv] MS. RAS C.1/1.2, 54-55.

[xxvi] MS. RAS C.1/1.2, 83-84.

[xxvii] MS. RAS C.1/1.2, 56-57; Kronk, *Cometography*, vol. 1, pp. 489-90.

[xxviii] MS. RAS W.3/1.12, 6-10.

[xxix] MS. RAS C.1/1.2, 60-70; Kronk, *Cometography*, vol. 1, pp. 491-93.

[xxx] MS. RAS W.3/1.12, 10.

[xxxi] MS. RAS C.1/1.2, 72-73; Kronk, *Cometography*, vol. 1, pp. 493-94.

[xxxii] MS. RAS W.3/1.12, 10-11; William Herschel, 'Miscellaneous Observations. Account of a Comet', *Philosophical Transactions of the Royal Society*, vol. 82 (1792), pp. 23-24.

[xxxiii] MS. RAS C.1/1.2, 79-80; MS. RAS W.3/1.12, 11; Caroline Herschel, 'An Account of the Discovery of a Comet', *Philosophical Transactions of the Royal Society*, vol. 84 (1794), p. 1.

[xxxiv] MS. RAS C.1/1.2, 90; MS. RAS W.3/1.12, 12.

[xxxv] Caroline Herschel and William Herschel, 'Account of the Discover of a New Comet. By Miss Caroline Herschel', *Philosophical Transactions of the Royal Society*, vol. 86 (1796), pp. 131-34.

---

[xxxvi] Kronk, *Cometography*, vol. 1, pp. 498-99.

[xxxvii] MS. RAS C.1/1.2, 91-94; MS. RAS W.3/1.12, 14; Kronk, *Cometography*, vol. 1, pp. 501-2.

[xxxviii] MS. RAS C.1/1.3, 99, 101, respectively.

[xxxix] MS. RAS C.1/1.3, 100-102; Kronk, *Cometography*, vol. 2, pp. 9-10.

[xl] MS. RAS C.1/1.3,102-4; MS. RAS W.3/1.12, 17-28; Kronk, *Cometography*, vol. 2, pp. 10-15.

[xli] MS. RAS C.1/1.3,104-106; MS. RAS W.3/1.12.1-7, 28-38, and W.3/1/14, 1-7; Kronk, *Cometography*, vol. 2, pp. 19-28.

[xlii] See Roberta J. M. Olson, Roberta and Jay M. Pasachoff, 'The 1816 Solar Eclipse and Comet 1811 I in John Linnell's Astronomical Album', *Journal for the History of Astronomy*, vol. 23 (1992), pp. 121-33; Olson and Paschoff, *Fire in the Sky*, pp. 121-28.

[xliii] MS. RAS C.1/1.3, 111; Kronk, *Cometography*, vol. 2, pp. 62-66.

[xliv] MS. RAS C.1/1.1, 12, 18. See Hoskin, *Discoverers of the Universe*, pp. 142-43.